\documentclass[12pt]{article}
\input{epsf}
\begin{document}
\def\hatt{{\hat t}}
\def\hatx{{\hat x}}
\def\hatth{{\hat \theta}}
\def\hatta{{\hat \tau}}
\def\hatrh{{\hat \rho}}
\def\hatva{{\hat \varphi}}
\def\gsim{\mathrel{\rlap{\lower4pt\hbox{\hskip1pt$\sim$}}
    \raise1pt\hbox{$>$}}}
\def\lsim{\mathrel{\rlap{\lower4pt\hbox{\hskip1pt$\sim$}}
    \raise1pt\hbox{$<$}}}
\def\p{\partial}
\def\nn{\nonumber}
\def\tils{{\tilde s}}
\def\tila{{\tilde a}}
\def\bart{{\bar t}}
\def\barx{{\bar x}}
\def\barh{{\bar \rho}}
\def\np#1#2#3{Nucl. Phys. {\bf B#1} (#2) #3}
\def\pl#1#2#3{Phys. Lett. {\bf B#1} (#2) #3}
\def\prl#1#2#3{Phys. Rev. Lett.{\bf #1} (#2) #3}
\def\pr#1#2#3{Phys. Rev. {\bf D#1} (#2) #3}
\def\ap#1#2#3{Ann. Phys. {\bf #1} (#2) #3}
\def\prep#1#2#3{Phys. Rep. {\bf #1} (#2) #3}
\def\rmp#1#2#3{Rev. Mod. Phys. {\bf #1}}
\def\cmp#1#2#3{Comm. Math. Phys. {\bf #1} (#2) #3}
\def\mpl#1#2#3{Mod. Phys. Lett. {\bf #1} (#2) #3}
\def\cqg#1#2#3{Class. Quant. Grav. {\bf #1} (#2) #3}
\def\jmp#1#2#3{J. Math. Phys. {\bf #1} (#2) #3}
\begin{titlepage}
\bigskip
\rightline{}
\rightline{hep-th/0207252}
\bigskip\bigskip\bigskip\bigskip
   \centerline{\Large \bf {Supersymmetries and Hopf-duality}}
   \bigskip
\centerline{\Large \bf {in the Penrose Limit of $AdS_3\times S^3\times T^4$}}
    \bigskip\bigskip
     \bigskip\bigskip

\centerline{\bf Jin-Ho Cho${}^*$, Taejin Lee${}^\dagger$ and Soonkeon Nam${}^\ddagger$}
\bigskip\bigskip
   \centerline{\em ${}^*$BK21 Physics Research Division and Institute of Basic Science}
\centerline{\em Sungkyunkwan University, 
Suwon 440-746, Korea}
\bigskip
   \centerline{\em ${}^\dagger$Department of Physics, Kangwon National University, 
 Chuncheon 200-701, Korea}
\centerline{\em \& Asia Pacific Center for Theoretical Physics, Pohang 790-784, Korea}
\bigskip
   \centerline{\em ${}^\ddagger$Department of Physics and Research Institute for Basic Sciences}
\centerline{\em Kyung Hee University, 
 Seoul 130-701, Korea}
\bigskip
\centerline{\sf jhcho@taegeug.skku.ac.kr, taejin@cc.kangwon.ac.kr, nam@khu.ac.kr}
    \bigskip\bigskip

\begin{abstract}
We investigate various aspects of the plane wave geometries obtained from D1/D5-brane system.
We study the effect of Hopf-duality on the supersymmetries preserved by the Penrose limit of $AdS_3\times S^3\times T^4$ geometry. In type-IIB case, we first show that the Penrose limit makes the size of the `would-be' internal torus comparable to that of the other directions. Based on this observation, we consider, in taking the Penrose limit, the generalization of the null geodesic to incorporate the tilted direction between the equator of $S^3$ and one of the torus directions. For generic values of the tilting angle, supersymmetries are not preserved. When the limit is taken along the torus direction, 16 supersymmetries are preserved. For the ordinary Penrose limit, 16 generic and 8 `supernumerary' supersymmetries are observed. In the Penrose limit of Hopf-dualized type-IIA geometry, only 4 supersymmetries are preserved. We classify all the Killing spinors according to their periodic properties along some relevant coordinates.
\end{abstract}
    \end{titlepage}

\section{Introduction}

As a testing ground for the truly stringy realization of AdS/CFT duality \cite{maldacena}, the plane wave geometry recently stood in the spotlight. It has appeared in the supergravity context as one of the maximally supersymmetric solutions for $D=11$, ${\cal N}=1$ supergravity \cite{kowalski}. Actually the plane wave geometry is contained in any spacetime as the tangent geometry to an arbitrary null geodesic \cite{penrose}. By taking the so-called Penrose limit to get this tangent space, one can reach the plane wave geometries from any geometries of $D=10,\,11$ supergravity theories \cite{gueven1,gueven2}, especially from any near-horizon AdS geometries of type IIB superstring theories and M-theory \cite{blau2,blau3}. One important property in these plane wave geometries is that they preserve maximal number of supersymmetries, 32 (i.e., the same number of supersymmetries as that of the AdS geometries) \cite{blau1,figueroa}. 

Composite D-branes have more exotic property concerning the supersymmetries; the Penrose limit of their near-horizon geometries allow more supersymmetries than their AdS partners do \cite{cvetic,gauntlett}. As for the D3/D3-brane intersection, there are 16 supersymmetries in its near horizon geometry, i.e., $AdS_3\times S^3\times T^4$. In the Penrose limit, 8 extra supersymmetries, coined as `supernumerary' ones in Ref. \cite{cvetic}, appear. Related issues were discussed in Refs. \cite{lu2,singh,kumar}.

The purpose of this paper is to investigate various aspects of plane wave geometries obtained from D1/D5-brane system. We first consider the general cases where null geodesic involves the torus direction as well as the sphere direction in taking the Penrose limit. We then examine the supersymmetry properties of the resulting geometry. Next, we study the behavior of the supersymmetries (including the supernumerary ones) under the Hopf-duality acting on the plane wave limit of $AdS_3\times S^3\times T^4$ geometry\footnote{T-duality commutes with the Penrose limit \cite{gueven2}. The Hopf-duality is a T-duality along the Hopf fiber direction of odd dimensional sphere \cite{duff2}. In the case at hand, it is a nontrivial duality which results in $AdS_3$ geometry in type-IIA theory \cite{duff}.}.  All the Killing spinors obtained above are classified according to their periodic properties along those coordinates which inherit periodicity from $AdS_3\times S^3\times T^4$ geometry. We shall work in the context of low energy effective action of type-II superstring theories.

The size of torus, $T^4$ is usually considered negligibly small compared to the other factors of the geometry. Therefore all the Kaluza-Klein momentum modes except the zero mode are suppressed along the torus directions due to large mass gap. However in the Penrose limit, the torus size becomes comparable to that of the other factors of the geometry. This is because the Penrose limit focuses on small region just around a geodesic followed by null rays \cite{penrose}. As we will see explicitly in Section \ref{ii}, the first leading order of the metric is of order $1/R^2$ in the limit. We cannot {\it \`{a} priori} exclude higher Kaluza Klein modes around the torus in such a limit. These modes could provide another kind of R-charge in the dual CFT. Hence it makes sense to tilt the direction of null rays to incorporate the torus direction in taking the Penrose limit. (Conventionally the direction of null rays is taken along the equator direction of the $S^3$.)

It was shown in Ref. \cite{bergshoeff} that T-duality does preserve the Killing spinors when they are independent of the dual coordinate. In their argument, the dimensional reduction was essential. T-duality rules are simplified in the case of dimensional reduction; the supersymmetry transformation equations can be recast in a manifestly T-duality invariant form in terms of low dimensional theory (see \cite{bergshoeff} for details). In this setting, the condition for the Killing spinors becomes coincident with the condition for the supersymmetries to be preserved by the dimensional reduction; the Killing spinor survives dimensional reduction when its Lie derivative in the Killing direction used for the dimensional reduction vanishes. 

In general, the dimensional reduction is an easy way to achieve the isometry that is necessary for T-duality. However in the Hopf-duality, the fiber direction is already a Killing direction. We need not assume small fiber size to write the metric in the Kaluza-Klein style. The 3-sphere metric in the new coordinates is completely of Kaluza-Klein style, where Kaluza-Klein gauge field does not depend on the fiber coordinate (see Appendix \ref{hopf}). The question we may raise at this point is that what happens to the Killing spinors which depend on the coordinate associated with the isometry used for Hopf-duality. Some Killing spinors which depend on the dual coordinate may survive duality \cite{gueven2}. However, in this paper we show that this is not the case. All the Killing spinors which survive duality do not depend on the dual coordinate.

The coordinates of plane wave geometries arising in the Penrose limit inherit their periodic properties from the original coordinates of the near horizon geometry. This is in contrast with the situation of the plane wave solution obtained in Ref. \cite{kowalski} where the Penrose limit was not used at all. In our case, the Killing spinors should respect the periodicity of some relevant coordinates. This condition may eliminate some of the solutions satisfying the Killing spinor equations. We show in detail which spinors survive this condition. 

The paper is composed as follows. In Section \ref{ii}, we define the Penrose limit focusing on the region around the tilted null geodesic. For an arbitrary tilting angle $\alpha$, the geometry becomes that of plane wave. In Section \ref{iii}, we obtain explicitly the Killing spinors for type-IIB geometry discussed in Section \ref{ii}. In Section \ref{iv}, we perform Hopf-duality on the configuration of D1/D5-brane system. The near horizon geometry in the resulting type-IIA configuration is $AdS_3\times S^2\times T^5$. In Section \ref{v}, we obtain the Killing spinors for the Penrose limit of type-IIA geometry obtained in Section \ref{iv}. The Killing spinor equations are more involved due to the presence of several different kinds of fields. In Section \ref{vi}, we examine whether the obtained Killing spinors are compatible with the periodicity of some coordinates. 
In Section \ref{vii}, we discuss the origin of the supernumerary supersymmetries and the effect of the Hopf duality. The paper is concluded with some remarks on future works.

In appendices we collect some useful formulas relevant in this paper. Appendix \ref{a} summarizes the Killing spinor equations of type-IIB supergravity discussed in Refs. \cite{schwarz,gsw}. In Appendix \ref{b}, we obtain the explicit expressions for the Killing spinors in the $AdS_3\times S^3\times T^4$ background, written in global coordinates. Appendix \ref{iia} contains the Killing spinor equations of type-IIA supergravity. Some old but relevant literatures on these are Ref. \cite{huq,giani}. See also Ref. \cite{hassan} for more modern format. In Appendix \ref{hopf}, we show explicitly how to obtain the Hopf-fibration of three sphere.

Our main results are as follows:
For generic values of the tilting angle, supersymmetries are not preserved. If the limit is taken along the torus direction, 16 supersymmetries are preserved. For the ordinary Penrose limit, i.e., when the tilting angle $\alpha=0$, 16 generic and 8 `supernumerary' supersymmetries are observed. In the Penrose limit of Hopf-dualized type-IIA geometry, only 4 supersymmetries are preserved. We classify all the Killing spinors according to their periodic properties along the periodic coordinates.

\section{Penrose limit of D$1$/D$5$-brane near horizon geometry}\label{ii}

We begin with D1/D5-brane system:
\begin{eqnarray}\label{d1/d5}
ds^2=(H_1H_5)^{-\frac{1}{2}}\left(-dt^2+dx_5^2\right)+H_1^{\frac{1}{2}}H_5^{-\frac{1}{2}}\sum\limits_{i=6}^{9}dx_i^2+(H_1H_5)^{\frac{1}{2}}\sum\limits_{i=1}^{4}dx_i^2.
\end{eqnarray}
Here the harmonic functions $H_1$ and $H_5$ in four transverse directions $(r^2=\sum\limits_{i=1}^{4}x_i^2)$ are given as follows:
\begin{eqnarray}
H_1=1+\frac{g\alpha'N_1}{vr^2},\quad H_5=1+\frac{g\alpha'N_5}{r^2}.
\end{eqnarray}
We keep $v=(\prod\limits_{i=6}^{9}R_i)/\alpha'^{2}$ finite, with $x_i\sim x_i+2\pi R_i, (i=6,\cdots, 9)$, hence consider  $N_5$ of D$5$-branes wrapping around four-torus of the string size. The system produces the dilaton field $e^{-2\phi}=H_1^{-1}H_5$ and the Ramond-Ramond (RR) three form field strength $F^{(3)}=-\ast dH_5-dH_1^{-1}\wedge dt\wedge dx_5$. ($\ast$ denotes the Hodge dual in the transverse flat spatial four dimensions.)

We now consider the near horizon limit keeping $\alpha'$ finite.
In the region $r^2\ll g\alpha'N_1/v,\,\,g\alpha'N_5$,we have
\begin{eqnarray}
H_1\sim\frac{g\alpha'N_1}{vr^2},\quad H_5\sim\frac{g\alpha'N_5}{r^2}.
\end{eqnarray}
For the lengths $g\alpha'N_1/v,\,\,g\alpha'N_5$ to be comparable with each other, $N_1/v$ and $N_5$ should be of the same order. Let us denote their finite ratio, the `near horizon' value of the string coupling $e^\phi$, as
\begin{eqnarray}
\zeta=\sqrt{\frac{N_1}{vN_5}}. 
\end{eqnarray}
It is well-known that the geometry factorizes as $AdS_3\times S^3\times T^4$ in this near horizon region;
\begin{eqnarray}
ds^2=\frac{r^2}{R^2}\left(-dt^2+dx_5^2\right)+\frac{R^2}{r^2}dr^2+R^2d\Omega_3^2+\zeta\sum\limits_{i=6}^{9}dx_i^2.
\end{eqnarray}
The sizes of $AdS_3$ and $S^3$ are equal and are given by $R^2=g\alpha'\sqrt{\frac{N_1N_5}{v}}$.
With the extension of the $AdS_3$ part using the global coordinates, one obtains the metric,
\begin{eqnarray}\label{adsmetric}
\frac{ds^2}{R^2}=&&-\cosh^2\rho\,\, dt^2+d\rho^2+\sinh^2\rho \,\,d\varphi^2\nonumber\\
&&+\cos^2{\theta}\,\,d\psi^2+d\theta^2+\sin^2{\theta}\,\,d\chi^2+\frac{\zeta}{R^2}\sum\limits_{i=6}^{9}dx_i^2.
\end{eqnarray}
The RR three form field strength becomes
\begin{eqnarray}
F^{(3)}=2\,g\alpha'N_5\left(\cosh{\rho}\,\,\sinh{\rho}\,\,dt\wedge d\rho\wedge d\varphi+\cos{\theta}\,\,\sin{\theta}\,\,d\psi\wedge d\theta\wedge d\chi\right).
\end{eqnarray}

Now we take the Penrose limit.
With the hindsight that we cannot neglect the size of the torus, we consider tilted null geodesic:
\begin{eqnarray}
\left(
\begin{array}{c}
\tilde{\psi}\cr
\tilde{x}_6/R
\end{array}
\right)=\left(\begin{array}{rr}
\cos{\alpha}&-\sin{\alpha}\cr \sin{\alpha}&\cos{\alpha}
\end{array}
\right)
\left(
\begin{array}{c}
\psi\cr x_6/R
\end{array}
\right).
\end{eqnarray}
The Penrose Limit is defined as follows. We first define new coordinates;
\begin{eqnarray}\label{pplimit}
t=x^++\frac{x^-}{R^2},\quad \tilde{\psi}=x^+-\frac{x^-}{R^2},\quad \rho=\frac{z}{R},\quad \theta=\frac{y}{R},
\end{eqnarray}
and take the large $R$ limit keeping $l^2\equiv R^2(\zeta-1)$ finite. Then the leading order terms of the metric becomes\footnote{In the conventional Penrose limit, all the transverse coordinates (to the lightcone coordinates) are scaled uniformly, thus all the topological features are sent to infinity. However here, the torus part is intact about the scalings $\rho=z/R,\, \theta=y/R$, and keeps its topological feature. In this sense, the limit we are taking is not just a Penrose limit, rather a double scaling limit.}
\begin{eqnarray}\label{metric}
ds^2=&&-4dx^+dx^--\left(z^2+y^2\cos^2{\alpha}+l^2\sin^2{\alpha}\right)(dx^+)^2\nonumber\\
&&+dz^2+dy^2+z^2d\varphi^2+y^2d\chi^2+\left(\sin^2{\alpha}+\zeta\cos^2{\alpha}\right)d\tilde{x}_6^2+\zeta\sum\limits_{i=7}^9 dx_i^2
\end{eqnarray}
The RR three form field strength becomes
\begin{eqnarray}\label{F}
F^{(3)}=2\,g\alpha'N_5\left(\frac{z}{R^2}dx^+\wedge dz\wedge d\varphi +\cos{\alpha}\frac{y}{R^2}dx^+\wedge dy\wedge d\chi\right).
\end{eqnarray}
As we see above, the size of the torus is comparable to the other parts of the geometry. We cannot neglect the Kaluza-Klein modes, which justifies our proposal to take the angular momentum direction tilted.
Hereafter, for convenience, we abbreviate the component $g_{++}$ to $-{\cal H}$ and introduce Cartesian coordinates, $\bar{x}_i (i=2,\cdots,9)$ as 
\begin{eqnarray}
&&dz^2+z^2d\varphi^2\equiv d\bar{x}_2^2+d\bar{x}_3^2,\qquad
dy^2+y^2d\chi^2\equiv d\bar{x}_4^2+d\bar{x}_5^2, \nonumber\\
&&\left(\sin^2{\alpha}+\zeta\cos^2{\alpha}\right)d\tilde{x}_6^2\equiv d\bar{x}_6^2,\qquad
\zeta dx_i^2\equiv d\bar{x}_i^2,\,\,(i=7,\,8,\,9). 
\end{eqnarray}
We may set in general, ${\cal H}=\sum\limits^5_{i=2}A_{ij}\bar{x}^i\bar{x}^j+\mbox{constant}$.

\section{Supersymmetries in type-IIB case}\label{iii}

In order to count the number of unbroken supersymmetries, we need to solve the Killing spinor equations.
In vanishing fermion backgrounds, the Killing spinor equations (in the Einstein frame) of type-IIB supergravity are summarized as \cite{schwarz}
\begin{eqnarray}\label{killing}
&&\delta\lambda=-\frac{i}{24}\Gamma^{KLN}G_{KLN}\,\,\epsilon=0,\nonumber\\
&&\delta\psi_M={\cal D}_M\epsilon+\frac{1}{96}\left(\Gamma_M{}^{KLN}G_{KLN}-9\Gamma^{LN}G_{MLN}\right)\epsilon^*=0,
\end{eqnarray}
where the Roman indices $K,\,L,\cdots$ are pertinent to ten dimensions. The $SU(1,1)$-invariant 3-form fields,
\begin{eqnarray}
G=2i\zeta^{-\frac{1}{2}}\left(dx^+\wedge d\bar{x}^2 \wedge d\bar{x}^3+\cos{\alpha}\,\,dx^+\wedge d\bar{x}^4 \wedge d\bar{x}^5\right),
\end{eqnarray}
are constructed from the RR 3-form field strengths of Eq. (\ref{F}) and the dilaton field $\phi$. Details are given in Appendix \ref{a}.
The orthonormal frame is  
\begin{eqnarray}
&&e^{(+)}=\zeta^{-\frac{1}{4}}dx^+, \qquad e^{(-)}=\zeta^{-\frac{1}{4}}\left(2dx^-+\frac{{\cal H}}{2}dx^+\right),\nonumber\\
&&e^{(i)}=\zeta^{-\frac{1}{4}}d\bar{x}^i,\quad (i=2,\cdots,9)
\end{eqnarray}
with respect to which gamma matrices in Eq. (\ref{killing}), written in the coordinate frame, are recast as
\begin{eqnarray}\label{gdown}
&&\Gamma_+=\zeta^{-\frac{1}{4}}\left(\Gamma_{(+)}+\frac{{\cal H}}{2}\Gamma_{(-)}\right),\quad
\Gamma_-=2\zeta^{-\frac{1}{4}}\Gamma_{(-)},\nonumber\\
&&\Gamma_{i}=\zeta^{-\frac{1}{4}}\Gamma_{(i)},\,\,(i=2,\cdots, 9).
\end{eqnarray}
The upper indexed gamma matrices are
\begin{eqnarray}\label{gup}
\Gamma^+=\zeta^{\frac{1}{4}}\Gamma^{(+)}, \,
\Gamma^-=\zeta^{\frac{1}{4}}\left(-\frac{{\cal H}}{4}\Gamma^{(+)}+\frac{1}{2}\Gamma^{(-)}\right), \, \Gamma^i=\zeta^{\frac{1}{4}}\Gamma^{(i)}, \,\,(i=2,\cdots, 9).
\end{eqnarray}
The covariant derivatives ${\cal D}_M$ in the plane wave background (\ref{metric}) are
\begin{eqnarray}
{\cal D}_+=\partial_++\frac{1}{4}\partial_i {\cal H}\,\Gamma_{(-)(i)}, \quad{\cal D}_-=\partial_-, \quad{\cal D}_i=\partial_i, \,\,(i=2,\cdots, 9).
\end{eqnarray}

In the orthonormal frame, the Killing spinor equations become
\begin{eqnarray}\label{killing2}
&&\delta\lambda=\frac{\zeta^{\frac{1}{4}}}{2}\left(\Gamma^{(+)(2)(3)}+\cos{\alpha}\,\Gamma^{(+)(4)(5)}\right)\epsilon=0,\nonumber\\
&&\delta\psi_+=\partial_+\epsilon-\frac{1}{4}\partial_i{\cal H}\,\Gamma^{(+)(i)}\epsilon-\frac{i}{8}\left(\Gamma^{(-)}\Gamma^{(+)}+4\right)\left(\Gamma^{(2)(3)}+\cos{\alpha}\Gamma^{(4)(5)}\right)\epsilon^*=0,\nonumber\\
&&\delta\psi_-=\partial_-\epsilon=0,\nonumber\\
&&\delta\psi_i=\partial_i\epsilon+i\Omega_i\epsilon^*=0,\qquad (i=2,\cdots,9).
\end{eqnarray}
In the last equation, the matrices $\Omega_i$ are given by
\begin{eqnarray}
&&\Omega_2\equiv\frac{1}{8}\Gamma^{(+)(3)}\left(3+\cos{\alpha}\,\Gamma^{(2)(3)(4)(5)}\right)=\frac{1}{2}\Gamma^{(+)(3)},\nonumber\\
&&\Omega_3\equiv-\frac{1}{8}\Gamma^{(+)(2)}\left(3+\cos{\alpha}\,\Gamma^{(2)(3)(4)(5)}\right)=-\frac{1}{2}\Gamma^{(+)(2)},\nonumber\\
&&\Omega_4\equiv\frac{1}{8}\Gamma^{(+)(5)}\left(3\cos{\alpha}+\Gamma^{(2)(3)(4)(5)}\right)=\frac{1}{2}\Gamma^{(+)(5)}\cos{\alpha},\nonumber\\
&&\Omega_5\equiv-\frac{1}{8}\Gamma^{(+)(4)}\left(3\cos{\alpha}+\Gamma^{(2)(3)(4)(5)}\right)=-\frac{1}{2}\Gamma^{(+)(4)}\cos{\alpha}\nonumber,\\
&&\Omega_i=0,\qquad i=6,\cdots,9,
\end{eqnarray}
which were simplified by using the dilatino condition $\delta\lambda=0$.
The complex Weyl spinor $\epsilon$ can be written in terms of two Majorana-Weyl spinors as $\epsilon=\epsilon_1+i\epsilon_2$. In ten dimensions, one can use purely real representation of the Clifford algebra, which results in 
\begin{eqnarray}
\partial_j\partial_i\epsilon=-i\Omega_i\partial_j\epsilon^*=\Omega_i\Omega_j^*\epsilon=\Omega_i\Omega_j\epsilon=0 \qquad (i=2,\cdots 5).
\end{eqnarray}
We use $\delta\psi_i=0$ in the first equality, and  $\Gamma^{(+)}\Gamma^{(+)}=0$ in the last equality. Therefore, $\epsilon$ is linear in $\bar{x}^i,\quad (i=2,\cdots 5)$ and independent of $x^-,\,\bar{x}^i,\quad (i=6,\cdots 9)$;
\begin{eqnarray}\label{2bspinor}
\epsilon=\eta(x^+)-i\sum\limits_{i=2}^{5}\Omega_i\,\eta^*(x^+)\,\bar{x}^i.
\end{eqnarray}
Plugging this result into the first two equations of Eq. (\ref{killing2}), we have
\begin{eqnarray}\label{killing3}
&&\frac{\zeta^{\frac{1}{4}}}{2}\Gamma^{(+)}\left(\Gamma^{(2)(3)}+\cos{\alpha}\Gamma^{(4)(5)}\right)\otimes 1\!\!1\left(\matrix{\eta_1\cr \eta_2}\right)=0,\nonumber\\
&&\left(\matrix{\partial_+\eta_1\cr \partial_+\eta_2}
\right)-\frac{1}{2}\left(\Gamma^{(2)(3)}+\cos{\alpha}\Gamma^{(4)(5)}\right)\otimes \sigma_1\left(\matrix{\eta_1\cr \eta_2}
\right)=0,\\
&&\Omega_j\otimes \sigma_1\left(\matrix{\partial_+\eta_1\cr \partial_+\eta_2}\right)
+\left[\frac{1}{2}A_{ij}\Gamma^{(+)(i)}-\frac{1}{2}\left(\Gamma^{(2)(3)}+\cos{\alpha}\Gamma^{(4)(5)}\right)\Omega_j\right]\otimes 1\!\!1\left(\matrix{\eta_1\cr \eta_2}\right)=0,\nonumber
\end{eqnarray}
where the first equation comes from the dilatino condition and the second (third) equations are due to the terms of the gravitino equation, which are independent (dependent) of the coordinates $\bar{x}^i,\,i=2,\cdots 5$. The $2\times 2$ identity matrix $1\!\!1,$ and the Pauli matrix $\sigma_1$ act on the column index $I=1, 2$ of $\eta_I$. The second equation determines the dependence of $\eta_I(x^+)$ on the coordinate $x^+$ as
\begin{eqnarray}\label{spinor1}
\left(\matrix{\eta_1(x^+)\cr \eta_2(x^+)}\right)=\exp{\left[\frac{x^+}{2}\left(\Gamma^{(2)(3)}+\cos{\alpha}\Gamma^{(4)(5)}\right)\otimes \sigma_1\right]}\left(\matrix{\bar{\eta}_{1}\cr \bar{\eta}_{2}}\right),
\end{eqnarray} 
where $\bar{\eta}_{I},\,I=1,2$ are constant $16$-component Majorana-Weyl spinors. 
The other conditions in Eq. (\ref{killing3}) are recast into the conditions on the constant spinor $\bar{\eta}$ as
\begin{eqnarray}\label{constant}
&&\left(\Gamma^{(+)(2)(3)}+\cos{\alpha}\Gamma^{(+)(4)(5)}\right)\bar{\eta}_I=0,\nonumber\\
&&\left(\mu_i-1\right)\Gamma^{(+)(i)}\bar{\eta}_I=0,\quad (i=2, 3),\nonumber\\
&&\left(\mu_i-\cos^2{\alpha}\right)\Gamma^{(+)(i)}\bar{\eta}_I=0,\quad (i=4, 5),
\end{eqnarray}
where we set $A_{ij}=\mu_{i}\delta_{ij}$.

If $\Gamma^{(+)}\bar{\eta}_I=0$, all the Killing spinor equations are satisfied. This projects out half of the constant spinors.  Therefore when no coordinate is periodic, there are 16 unbroken supersymmetries in the generic case. From the other sixteen spinors $\bar{\eta}_1$ and $\bar{\eta}_2$ satisfying $\Gamma^{(-)}\bar{\eta}_I=0$, it is possible to obtain additional Killing spinors. The first equation of Eq. (\ref{constant}) reduces to $\Gamma^{(2)(3)(4)(5)}\bar{\eta}_I=\cos{\alpha}\,\,\bar{\eta}_I$ implying $\cos^2{\alpha}=1$. The remaining Killing spinor equations are satisfied if $\mu_i=1,\, (i=2,\cdots,5)$. This means that there are eight extra supersymmetries under the conditions $\mu_i=1,\, (i=2,\cdots,5)$. Note that the metric of Eq. (\ref{metric}) automatically satisfies the above conditions. Boosting the lightcone coordinates as $x^+\rightarrow x^+\mu,\,\, x^-\rightarrow x^-\mu^{-1}$ does not change the result because it does not only add extra factor $\mu^2$ to all $\mu_i,\,(i=2,\cdots 5)$, but also changes the RR field of Eq. (\ref{F}) by a factor $\mu$ so that Eq. (\ref{constant}) remains the same. Since some coordinates are periodic, we should call our special attention to the periodic properties of the Killing spinors. This will be discussed in Section \ref{vi}.

\section{Hopf-duality and $AdS_3$ in type-IIA supergravity}\label{iv}

Here we perform Hopf-duality on the D-brane configuration we considered in the previous sections, to obtain a type-IIA configuration\footnote{In Ref. \cite{duff}, this was done in six dimensional truncated theory. In this paper, we work in full ten dimensions for completeness and the procedure becomes even more simplified since the standard T-duality rules of ten dimensions are applicable.}.
Hopf duality is a T-duality along the $U(1)$ fiber direction of the odd dimensional sphere \cite{duff}. For the case at hand, it is based on the following Hopf fibration of a three-sphere: 
\begin{eqnarray}
d\Omega_3^2=\frac{1}{4}d\bar{\Omega}_2^2+\frac{1}{4\alpha'}\left(l_s\,d\bar{\chi}+l_s\bar{{\cal A}}\right)^2,\, (l_s^2=\alpha'), {\rm where} \ d\bar{\Omega}_2^2 =d\bar{\theta}^2 + 
\cos ^2\bar{\theta} d\bar{\psi}^2.
\end{eqnarray} 
The field strength ${\cal F}$ of the Kaluza-Klein gauge field ${\cal A}\equiv l_s\bar{{\cal A}}$ is given in terms of the K\"{a}hler two form $J=d\bar{\Omega}_2$, as  ${\cal F}=-l_sJ$. The explicit derivation is given in Appendix \ref{hopf}.

Taking T-duality along $\bar{\chi}$-direction on the geometry (\ref{d1/d5}), we get the following string metric.
\begin{eqnarray}
ds_{st}^2&=&(H_1H_5)^{-\frac{1}{2}}\left(-dt^2+dx_5^2\right)+H_1^{\frac{1}{2}}H_5^{-\frac{1}{2}}\sum\limits_{i=6}^{9}dx_i^2\nonumber\\
&+&(H_1H_5)^{\frac{1}{2}}\left(dr^2+\frac{r^2}{4}d\bar{\Omega}_2^2\right)+4(H_1H_5)^{-\frac{1}{2}}\frac{\alpha'^2}{r^2}d\bar{\chi}^2.
\end{eqnarray}
The RR 3-form field in type-IIB theory splits into a RR 2-form field $F^{(2)}$ and a RR 4-form field $F^{(4)}$:
\begin{eqnarray}
&&F^{(2)}=\frac{1}{4}gl_sN_5 \,d\bar{\Omega}_2,\nonumber\\
&&F^{(4)}=-2gl_s\alpha'\frac{N_1}{v}r^{-3}H_1^{-2}dt\wedge dr\wedge dx_5\wedge d\bar{\chi}.
\end{eqnarray}
The Kaluza-Klein gauge field along $\bar{\chi}$-direction becomes Neveu-Schwarz (NS) field to produce $H^{(3)}$. The dilaton field $\phi'$ in type-IIA theory was determined such that the invariant measure $\sqrt{-g_{st}}\,e^{-2\phi}$ of type-II string frame is invariant under the T-duality:
\begin{eqnarray}
&&H^{(3)}=\alpha'd\bar{\Omega}_2\wedge d\bar{\chi},\nonumber\\
&&e^{\phi'}=2H_1^{\frac{1}{4}}H_5^{-\frac{3}{4}}l_s/r.
\end{eqnarray}

In the near-horizon, the geometry becomes $AdS_3\times S^2\times T^5$:
\begin{eqnarray}
\frac{ds^2}{R^2}=&&-\cosh^2\rho\,\, dt^2+d\rho^2+\sinh^2\rho \,\,d\varphi^2\nonumber\\
&&+\frac{1}{4}\left(\cos^2{\bar{\theta}}\,\,d\bar{\psi}^2+d\bar{\theta}^2\right)+\frac{4\alpha'^2}{R^4}d\bar{\chi}^2+\frac{\zeta}{R^2}\sum\limits_{i=6}^{9}dx_i^2,
\end{eqnarray} 
where $-\pi/2\le\bar{\theta}\le \pi/2$. 
The other fields become in the near horizon limit
\begin{eqnarray}
&&F^{(2)}=-\frac{1}{4}gl_sN_5\cos{\bar{\theta}}\,d\bar{\psi} \wedge d\bar{\theta},\nonumber\\
&&F^{(4)}=-2gl_s\alpha'N_5\cosh{\rho}\,\,\sinh{\rho}\,\,dt\wedge d\rho\wedge d\varphi \wedge d\bar{\chi},\nonumber\\
&&H^{(3)}=-\alpha'\cos{\bar{\theta}}\,d\bar{\psi} \wedge d\bar{\theta} \wedge d\bar{\chi},\nonumber\\
&&e^{\phi'}=2R/gl_sN_5.
\end{eqnarray}

\section{Supersymmetries in type-IIA case}\label{v}

We would like to see if the Penrose limit results in extra supersymmetries in type-IIA case also. 
Taking the Penrose limit as follows
\begin{eqnarray}\label{pplimit78}
&&t=x^++\frac{x^-}{R^2},\quad \frac{\bar{\psi}}{2}=x^+-\frac{x^-}{R^2},\nonumber\\
&&\rho=\frac{z}{R},\quad \frac{\bar{\theta}}{2}=\frac{\bar{x}_4}{R},\quad
\bar{\chi}= \frac{R\bar{x}_5}{2\alpha'},\quad \zeta^{1/2}x_i = \bar{x}_i.
\end{eqnarray}
we get the following plane wave geometry,
\begin{eqnarray}\label{2ametric}
ds^2=&&-4dx^+dx^--\left(\bar{x}_2^2+\bar{x}_3^2+4\bar{x}_4^2\right)(dx^+)^2+\sum\limits_{i=2}^9 d\bar{x}_i^2.
\end{eqnarray}
In the same limit the RR fields and the NS field behave as
\begin{eqnarray}\label{rrns}
&&F^{(2)}=-\frac{gl_sN_5}{R}dx^+\wedge d\bar{x}_4,\nonumber\\
&&F^{(4)}=-\frac{gl_sN_5}{R}dx^+\wedge d\bar{x}_2\wedge d\bar{x}_3\wedge d\bar{x}_5,\\
&&H^{(3)}=-2dx^+\wedge d\bar{x}_4 \wedge d\bar{x}_5.\nonumber
\end{eqnarray}
It seems that only the NS field $H^{(3)}$ survives in the large $R$ limit. However, as we will see below, these fields appear in the Killing spinor equation with appropriate dilaton factors so that all the fields survive in the large $R$ limit.
The main difference between the geometry of type-IIA theory and that of type-IIB theory is that the metric component $g_{++}=-{\cal H}$ is a function quadratic only in three coordinates $\bar{x}_2,\bar{x}_3,\bar{x}_4$.

In type-IIA supergravity, the bosonic part of the Killing spinor equations are summarized in Appendix \ref{iia} (see also Ref. \cite{giani,hassan}).
Here, all fermions are represented by Majorana spinors.
The relation of gamma matrices in the coordinate frame and those in the orthonormal Einstein frame is basically the same as Eqs. (\ref{gdown},\ref{gup}) except that $\zeta$ is replaced by the `near horizon' value of type-IIA string coupling $e^{\phi'}=2R/gl_sN_5$. With the explicit expressions of form fields (\ref{2ametric}) and (\ref{rrns}), the Killing spinor equations reduce to
\begin{eqnarray}\label{killing4}
\sqrt{2}\delta\lambda&=&-e^{\frac{\phi'}{4}}\left(\frac{1}{2}\Gamma^{(+)(4)(5)}-\frac{3}{8}\Gamma^{(+)(4)}+\frac{1}{8}\Gamma^{(+)(2)(3)(5)}\Gamma^{11}\right)\epsilon=0,\nonumber\\
\delta\psi_+&=&\partial_+\epsilon-\sum\limits_{i=2}^3 \bar{x}_i\Gamma^{(+)(i)}\epsilon-4\bar{x}_4\Gamma^{(+)(4)}\epsilon+\frac{1}{4}\Gamma^{(-)}\Gamma^{(+)}\left(2\Gamma^{(4)(5)}+\Gamma^{(4)}\right)\Gamma^{11}\epsilon\nonumber\\
&&+\frac{1}{2}\left(\Gamma^{(4)(5)}\Gamma^{11}-
\frac{1}{2}\Gamma^{(4)}\Gamma^{11}+\frac{1}{2}\Gamma^{(2)(3)(5)}\right)\epsilon=0,\nonumber\\
\delta\psi_i&=&\partial_i\epsilon-\Omega_i\epsilon=0 \qquad (i=2,\cdots 9),
\end{eqnarray}
where
\begin{eqnarray}\label{omega}
\Omega_i\equiv-\frac{\Gamma^{(i)}}{2}\left(a_i\Gamma^{(+)(4)(5)}+b_i\Gamma^{(+)(4)}\right)\Gamma^{11},
\end{eqnarray}
with $(a_2, b_2)=(a_3, b_3)=(1,1)$, $(a_4, b_4)=(0,-1)$, 
$(a_5, b_5)=(2,1)$, and $(a_6, b_6)=\cdots =(a_9, b_9)=(-1,-1/2)$. Here $\epsilon$ is a Majorana spinor composed of two Weyl spinors of opposite chiralities.

Since $\partial_j\partial_i\epsilon=\Omega_i\Omega_j\epsilon=0,\,(i,j=2,\cdots 9)$, the Killing spinor $\epsilon$ is at most linear in these coordinates; 
\begin{eqnarray}\label{sol0}
\epsilon(x^+,\bar{x}_i)=\eta(x^+)+\sum\limits_{i=2}^9\Omega_i\eta(x^+)\bar{x}_i.
\end{eqnarray}
Inserting this expression into the first and the second equations of Eq. (\ref{killing4}), we get
\begin{eqnarray}
&&\left(4\Gamma^{(+)(4)(5)}-3\Gamma^{(+)(4)}+\Gamma^{(+)(2)(3)(5)}\Gamma^{11}\right)\eta(x^+)=0,\nonumber\\
&&\partial_+\eta(x^+)+\frac{1}{4}\Gamma^{(-)}\Gamma^{(+)}\left(2\Gamma^{(4)(5)}+\Gamma^{(4)}\right)\Gamma^{11}\eta(x^+)\nonumber\\
&&\qquad+\frac{1}{2}\left(\Gamma^{(4)(5)}\Gamma^{11}-\frac{1}{2}\Gamma^{(4)}\Gamma^{11}+\frac{1}{2}\Gamma^{(2)(3)(5)}\right)\eta(x^+)=0,\nonumber\\
&&\left(\frac{1}{2}\sum\limits_{i=2}^{9}\left(\Omega_i\left(2\Gamma^{(4)(5)}+\Gamma^{(4)}\right)\Gamma^{11}-\left[\Omega_i,\,\Gamma^{(4)(5)}\Gamma^{11}-\frac{1}{2}\Gamma^{(4)}\Gamma^{11}+\frac{1}{2}\Gamma^{(2)(3)(5)}\right]\right)\bar{x}_i\right.\nonumber\\
&&\qquad\left.-\sum\limits_{i=2}^{3}\Gamma^{(+)(i)}\bar{x}_i-
4\Gamma^{(+)(4)}\bar{x}_4\right)\eta(x^+)=0.
\end{eqnarray}
The second equation determines $x^+$-dependence of the Killing spinor; 
\begin{eqnarray}\label{sol1}
\eta(x^+)&=&\exp\left[{-\frac{x^+}{4}\Gamma^{(-)}\Gamma^{(+)}\left(2\Gamma^{(4)(5)}+\Gamma^{(4)}\right)\Gamma^{11}}\right.\nonumber\\
&&\qquad\left.{-\frac{x^+}{2}\left(\Gamma^{(4)(5)}\Gamma^{11}-\frac{1}{2}\Gamma^{(4)}\Gamma^{11}+\frac{1}{2}\Gamma^{(2)(3)(5)}\right)}\right]\bar{\eta},
\end{eqnarray}
where $\bar{\eta}$ is a constant spinor.
The third equation is specified as
\begin{eqnarray}\label{killing5}
&&\Gamma^{(i)(+)(5)}\eta(x^+)=0,\qquad (i=2,3),\nonumber\\
&&\frac{7}{2}\Gamma^{(4)(+)}\eta(x^+)=0,\nonumber\\
&&\Gamma^{(i)}\left(4\Gamma^{(+)(5)}-5\Gamma^{(+)}\right)\eta(x^+)=0,\qquad (i=5,\cdots, 9).
\end{eqnarray}
One can easily see that the same conditions apply to the constant spinors $\bar{\eta}$. Therefore only the constant spinors satisfying $\Gamma^{(+)}\bar{\eta}=0$ generate the solutions of the Killing spinor equations (\ref{killing4}). Due to this condition, the linear dependence of the solutions (\ref{sol0}) on the coordinates, $\bar{x}_i,\,(i=2,\cdots, 9)$ disappears. If the coordinate $x^+$ is not periodic, $16$ supersymmetries are preserved and there is no supernumerary supersymmetry. However, the coordinate $x^+$ inherits its periodicity from the coordinate $\bar{\psi}$, thus the solutions should respect the periodicity. In the next section we will elaborate on this point.   

\section{Periodicity of the Killing Spinors}\label{vi}

The coordinates of the plane wave geometries of Eqs. (\ref{metric}), (\ref{2ametric}) inherit their periodicities from the original coordinates of the near horizon geometry. Since the transition functions of the spin manifold are given in terms of the spinor representation of the Lorentz group $SO(9,1)$, the Killing spinor should be either periodic or anti-periodic along those compact directions (see p. 276 of Ref. \cite{gsw} for details). This may eliminate some of the solutions we have constructed so far.

\subsection{IIB Case}

Let us consider type-IIB case first. The periodic coordinates are
\begin{eqnarray}
&&\left(x^+,\,\,x^-\right)\sim\left(x^++\pi k\cos{\alpha}-\pi n\frac{R_6}{R}\sin{\alpha},\,\,\,\,x^--\pi k R^2\cos{\alpha}+\pi nRR_6 \sin{\alpha}\right),\nonumber\\
&&\bar{x}_6\sim \bar{x}_6+\left(\sin{\alpha}+\zeta\cos{\alpha}\right)^{\frac{1}{2}}\left(2\pi kR\sin{\alpha}+2\pi nR_6\cos{\alpha}\right),\nonumber\\
&&\bar{x}_i\sim\bar{x}_i+2\pi R_i\zeta^{\frac{1}{2}},\qquad (i=7,8,9)
\end{eqnarray}
where $k,n\in Z\!\!\!Z$. Their periodicities are compatible with `$\psi\sim\psi+2\pi k$' and `$x_6\sim x_6+2\pi n R_6$'. The Killing spinor $\epsilon$ in Eq. (\ref{2bspinor}) depends only on the periodic coordinate $x^+$. Then the (anti-)periodic condition reduces to 
\begin{eqnarray}
&&\exp{\left[\beta\left(\Gamma^{(2)(3)}+\cos{\alpha}\Gamma^{(4)(5)}\right)\otimes \sigma_1\right]}\left(\matrix{\bar{\eta}_{1}\cr \bar{\eta}_{2}}\right)=\pm\left(\matrix{\bar{\eta}_{1}\cr \bar{\eta}_{2}}\right),\nonumber\\
&&\qquad \beta\equiv \frac{\pi}{2}\lim_{R\rightarrow\infty}\left(k\cos{\alpha}-n\frac{R_6}{R}\sin{\alpha}\right)=\frac{\pi k}{2}\cos{\alpha}.
\end{eqnarray}
The upper/lower sign corresponds to the periodic/anti-periodic condition. The resulting Killing spinors are always periodic for the cycle $(k, n)=(0, 1)$. Expanding the exponent, we get
\begin{eqnarray}\label{expand}
&&\left[\sin{\beta}\,\,\Gamma^{(2)(3)}\otimes\sigma_1\pm\sin{\left(\beta \cos{\alpha}\right)}\Gamma^{(4)(5)}\otimes\sigma_1\right]\left(\matrix{\bar{\eta}_{1}\cr \bar{\eta}_{2}}\right)\nonumber\\
&&=\left[-\cos{\beta}\pm \cos{\left(\beta \cos{\alpha}\right)}\right]\left(\matrix{\bar{\eta}_{1}\cr \bar{\eta}_{2}}\right).
\end{eqnarray}
It follows from repetition of this eigenvalue equation
\begin{eqnarray}
\left[\sin{\beta}\,\sin{\left(\beta\cos{\alpha}\right)}\,\Gamma^{(2)(3)(4)(5)}+\cos{\beta}\,\cos{\left(\beta\cos{\alpha}\right)}\mp1\right]\left(\matrix{\bar{\eta}_{1}\cr \bar{\eta}_{2}}\right)=0,
\end{eqnarray}
which implies $\Gamma^{(2)(3)(4)(5)}\bar{\eta}_I=\xi\,\bar{\eta}_I$. The value of $\xi$ can be either $+1$ or $-1$.
Thus the condition satisfying Eq. (\ref{expand}) is summarized as
\begin{eqnarray}
\frac{k}{2}\left(1-\xi\cos{\alpha}\right)\cos{\alpha}
=\left\{
\begin{array}{llr}
2m & \quad\mbox{periodic case (P)},&\cr
&& \quad m\in Z\!\!\!Z\cr
2m+1 & \quad\mbox{anti-periodic case (A)}.&
\end{array}
\right.
\end{eqnarray}

For the unit cycle $(k, n)=(1, 0)$, the range of the cosine function allows $m=0$ only. When $\cos{\alpha}=0$ or $\cos{\alpha}=\xi$, only the periodic spinors are possible, while the anti-periodic spinors are possible when $\cos{\alpha}=-\xi$. 

Table 1. summarizes the result for type-IIB case: When $\cos^2{\alpha}=1$, i.e. when the Penrose limit is taken purely along the 3-sphere direction, there are $24$ Killing spinors. Eight of them are periodic for the cycle $(k, n)=(1, 0)$, satisfying $\Gamma^{(+)}\bar{\eta}_I=0$ and $\Gamma^{(2)(3)(4)(5)}\bar{\eta}_I=\cos{\alpha}\,\,\bar{\eta}_I$, another eight Killing spinors are anti-periodic for the cycle $(k, n)=(1, 0)$, satisfying $\Gamma^{(+)}\bar{\eta}_I=0$ and $\Gamma^{(2)(3)(4)(5)}\bar{\eta}_I=-\cos{\alpha}\,\,\bar{\eta}_I$. The other eight Killing spinors satisfying $\Gamma^{(-)}\bar{\eta}_I=0$ and $\Gamma^{(2)(3)(4)(5)}\bar{\eta}_I=\cos{\alpha}\,\,\bar{\eta}_I$ are `supernumerary' ones and do not depend on the coordinate $x^+$, thus are periodic for the cycles $(1, 0)$. When $\cos{\alpha}=0$, i.e., the Penrose limit is taken purely along one of the torus directions, only $16$ Killing spinors satisfying $\Gamma^{(+)}\bar{\eta}_I=0$ are possible. They are periodic for the cycle $(k, n)=(1, 0)$. As for the cycle $(k, n)=(0, 1)$, all the above Killing spinors are periodic. When $\cos^2{\alpha}\ne 0, \pm1$, no (anti-)periodic Killing spinor is possible. 

\begin{table}[h]
\begin{tabular}{||r|c|r|c|c|c|c||}
\hline\hline
$\cos{\alpha}$&$\Gamma^{(+)}$ or $\Gamma^{(-)}$&$\xi$&$(1,0)$&$(0,1)$&\multicolumn{2}{c||}{ number of Killing spinors}\\ \hline
 & $\Gamma^{(+)}\bar{\eta}=0$ & 1 &  P & P & 8 & \\ \cline{3-6}
 1 &                        &-1 & A & P & 8 & 24  \\ \cline{2-6}
  & $\Gamma^{(-)}\bar{\eta}=0$ & 1 & P & P & 8 &   \\ \cline{3-6}
  &                        &-1 &$\cdot$  & $\cdot$  &0 &\\ \hline\hline
0 & $\Gamma^{(+)}\bar{\eta}=0$ &  1 & P & P & 8 & 16\\ \cline{3-6}
  &                        & -1 & P & P & 8& \\ \hline\hline
& $\Gamma^{(+)}\bar{\eta}=0$ & 1 &  A & P & 8 & \\ \cline{3-6}
 -1 &                        &-1 &  P & P & 8 & 24  \\ \cline{2-6}
  & $\Gamma^{(-)}\bar{\eta}=0$ & -1 &  P & P & 8 &   \\ \cline{3-6}
  &                        &1 & $\cdot$  & $\cdot$  &0 &\\ \hline\hline
\end{tabular}
\caption{\footnotesize $\alpha$ is the mixing angle between the $S^3$-direction and the $T^4$-direction. $\Gamma^{(2)(3)(4)(5)}\bar{\eta}=\xi\,\,\bar{\eta}$.
$(1, 0)$ or $(0, 1)$ denotes the the cycle $(k, n)$. Character `P' denotes `periodic' while `A' denotes `anti-periodic'.}
\end{table}

\subsection{IIA case}

Now we find Killing spinors of type-IIA case which are compatible with the periodic coordinates; 
\begin{eqnarray}\label{2aperiod}
&&\left(x^+,\,\,x^-\right)\sim\left(x^++\pi/2,\,\,x^--\pi R^2/2\right),\nonumber\\
&&\bar{x}_5\sim \bar{x}_5+8\pi\frac{\alpha'}{R},\nonumber\\
&&\bar{x}_i\sim \bar{x}_i+2\pi R_i\zeta^\frac{1}{2}\qquad (i=6,\cdots 9).
\end{eqnarray}
The first equation corresponds to the periodic coordinate $\bar{\psi}\sim\bar{\psi}+2\pi$ and the second equation concerns the periodic fiber coordinate $\bar{\chi}\sim \bar{\chi}+4\pi$. As was discussed in Sec. \ref{v}, the linear dependence of the Killing spinors on the coordinates $\bar{x}_i,\,(i=2,\cdots,9)$ disappears when we impose the condition $\Gamma^{(+)}\bar{\eta}=0$. The (anti-)periodicity condition along the coordinate $x^+$, may be transcribed into the following condition for the constant spinors $\bar{\eta}$,
\begin{eqnarray}
\exp{\left[-\frac{\pi}{4}\left(\Gamma^{(4)(5)}\Gamma^{11}-\frac{1}{2}\Gamma^{(4)}\Gamma^{11}
+\frac{1}{2}\Gamma^{(2)(3)(5)}\right)\right]}\bar{\eta}=\pm\bar{\eta}.
\end{eqnarray}
Expanding the exponent, we get
\begin{eqnarray}\label{expo}
&&\left(1-\Gamma^{(4)(5)}\Gamma^{11}\right)\left[\cos{\frac{\pi}{4}}\left(1+\Gamma^{(2)(3)(4)(5)}\Gamma^{11}\right)\right.\nonumber\\
&&\qquad\left.+\cos^2{\frac{\pi}{4}}\left(\left(1-\Gamma^{(2)(3)(4)(5)}\Gamma^{11}\right)+\left(\Gamma^{(4)}\Gamma^{11}-\Gamma^{(2)(3)(5)}\right)\right)\right]\bar{\eta}
=\pm 2\bar{\eta}.
\end{eqnarray}

Noting that $\cos{\pi/4}=1/\sqrt{2}$ is an irrational number,
we may split the equation (\ref{expo}) into two parts, one with $\cos(\pi/4)$ 
and the other without;
\begin{eqnarray}
&&\left(1+\Gamma^{(2)(3)(4)(5)}\Gamma^{11}\right)\bar{\eta}=0,\label{con2}\\
&&\left(1-\Gamma^{(5)}-\Gamma^{(2)(3)}\left(1+\Gamma^{(5)}\right)\right)\bar{\eta}=\pm2\bar{\eta}\label{con3}.
\end{eqnarray}
The second equation (\ref{con3}) may be rewritten as 
\begin{eqnarray}
(1+ \Gamma^{(5)})(1+ \Gamma^{(2)(3)}) \,\,\bar{\eta} =
\left\{\begin{array}{ll}
   0 & \qquad \mbox{(P)}, \\
   4 \,\,\bar{\eta} & \qquad \mbox{(A)}.
\end{array} \right.
\end{eqnarray}
Note that since $(\Gamma^{(2)(3)})^2 = -1$, it cannot have real 
eigenvalues. It readily implies that in the anti-periodic case we
do not have a solution and in the periodic case we have one only when 
\begin{eqnarray}\label{con4}
\left(\Gamma^{(5)}+1\right)\bar{\eta}=0.
\end{eqnarray}
Recalling that the constant spinors are subject to the 
condition $\Gamma^{(+)}\bar{\eta}=0$, we find that Eqs. (\ref{con2}) 
and (\ref{con4}) yield 4 unbroken supersymmetries in this type-IIA case. 
All these 4 Killing spinors are periodic along $x^+$-directions.

\section{Discussions}\label{vii}

It seems appropriate to give some thought to the reason 
why `supernumerary' Killing spinors are generated in the Penrose limit of 
$AdS_3\times S^3\times T^4$. In Ref. \cite{cvetic}, these supernumerary 
supersymmetries were shown to be related to the linearly realized world-sheet 
supersymmetries of the lightcone superstrings on the corresponding background. 
Here we give a more direct and explicit explanation for the supernumerary 
supersymmetries as follows. By a simple coordinate transformation, we can 
relate the gamma matrices $\gamma^{(\mu)}$ based on the global coordinates 
of $AdS_3\times S^3$ (see Eq. (\ref{gammaup}) in Appendix \ref{b}) with the 
gamma matrices $\Gamma^{(\mu)}$ of (\ref{gup}) constructed for the 
corresponding plane wave geometry:
\begin{eqnarray}
\gamma^{(0)}&=&\left(R-\frac{\cal H}{4R}\right)\cosh{\rho}\Gamma^{(+)}+\frac{1}{2R}\cosh{\rho}\Gamma^{(-)},\nonumber\\
\gamma^{(1)}&=&\left(R+\frac{\cal H}{4R}\right)\cos{\theta}\Gamma^{(+)}-\frac{1}{2R}\cos{\theta}\Gamma^{(-)},\nonumber\\
\gamma^{(2)}&=&\cos{\varphi}\Gamma^{(2)}+\sin{\varphi}\Gamma^{(3)},\nonumber\\
\gamma^{(3)}&=&\frac{R\sinh{\rho}}{z}\left(-\sin{\varphi}\Gamma^{(2)}+\cos{\varphi}\Gamma^{(3)}\right),\nonumber\\
\gamma^{(4)}&=&\cos{\chi}\Gamma^{(4)}+\sin{\chi}\Gamma^{(5)},\nonumber\\
\gamma^{(5)}&=&\frac{R\sin{\theta}}{y}\left(-\sin{\chi}\Gamma^{(4)}+\cos{\chi}\Gamma^{(5)}\right).
\end{eqnarray} 
Note that in the Penrose limit (\ref{pplimit}), both $\gamma^{(0)}$ and 
$\gamma^{(1)}$ coalesce to a projector, $R\,\,\Gamma^{(+)}$ and the dilatino 
condition (\ref{adsdilatino}) changes as follows:
\begin{eqnarray}
&&\frac{1}{2}R^{-1}\zeta^{\frac{1}{4}}\left(\gamma^{(0)(2)(3)}+\gamma^{(1)(4)(5)}\right)\epsilon=0\nonumber\\
&&\Longrightarrow \frac{1}{2}\zeta^{\frac{1}{4}}\Gamma^{(+)}\left(\Gamma^{(2)(3)}+\Gamma^{(4)(5)}\right)\epsilon=0.
\end{eqnarray}
As a consequence, the operator becomes factorized into $\Gamma^{(+)}$ and $\left(\Gamma^{(2)(3)}+\Gamma^{(4)(5)}\right)$. Each of them can annihilate 
the spinor $\epsilon$ separately. This opens up a new possibility 
to obtain additional Killing spinors.

Let us make a few remarks on the periodicity analysis carried out in this paper. Throughout the analysis, Penrose limits with generic values of tilting angle ($\cos{\alpha}\ne 0,\pm 1$) turned out to give nonsupersymmetric plane wave geometries. The same analysis also excluded 12 aperiodic spinors out of 16 Killing spinors of the Hopf-dualized geometry. These results seem to be in contradiction with the assertion made in Ref. \cite{blau3}: the number of supersymmetries of a supergravity background never decreases in the Penrose limit. A few words on this `seeming' contradiction are in order. First, the paper \cite{blau3} did only consider local properties of supergravity configuration and not the global properties including the periodicity of the Killing spinors. Before the periodicity analysis, our results were perfectly in agreement with their assertion. Second, the periodicity of the Killing spinors is not a hereditary property in their sense. This can be easily understood by looking at the Killing spinors (\ref{adsKilling}) of the original $AdS$ geometry. Those Killing spinors are all {\it anti-periodic} along several periodic coordinates. However, most coordinate dependency (except $x^+$-dependency) disappears in the Penrose limit making the spinors {\it periodic} along those coordinates. The reason why we have to look into the periodicity of the Killing spinors is as follows. As is also said in Ref. \cite{blau3}, one can always identify the points between $AdS$ geometry (\ref{adsmetric}) and the plane wave geometry (\ref{metric}). A periodic array of points in the $AdS$ geometry remain still as a periodic array in the Penrose limit, although the period becomes rescaled in general. This implies that the periodic property of the geometry is hereditary although the periodicity of the Killing spinors is not. The spinor fields over the geometry should definitely respect these periodic properties of the geometry.

Let us now discuss the effect of Hopf-duality on the Killing spinors. 
As we observed before, the standard Killing spinors depend at most on $x^+$ due to the condition $\Gamma^{(+)}\bar{\eta}= 0$, both in type-IIB and type-IIA cases. Especially, the periodic Killing spinors are constant ones. Since $x^+$ looks independent of the fiber direction $\chi$ along which the T-duality is taken, we are apt to conclude that the standard Killing spinors may be kept intact by Hopf duality\footnote{the supernumerary Killing spinors of type-IIB case depend only on the transverse coordinates $\bar{x}^i,\,\,(i=2,\cdots,5)$, with which the fiber direction is concerned.}. 
However, this is misleading. There are two important points we overlook easily.
First, the fiber coordinate is not $\chi$ but $\bar{\chi}$ as we see in Eq. (\ref{fiber}) of Appendix \ref{hopf}. Making use of Eqs. (\ref{pplimit}), (\ref{pplimit78}), and (\ref{new}), one obtains the following relation,
\begin{eqnarray}
x^+_B=\frac{1}{2}\left(t+\psi\right)
=\frac{1}{2}\left(t+\frac{1}{2}\left(\bar{\psi}+\bar{\chi}\right)\right)
=x^+_A+\frac{1}{4}\bar{\chi},
\end{eqnarray}
where we set $\alpha=0$ in type-IIB case. The subscripts $A, B$ distinguish between the lightcone coordinates $x^+$'s in type-IIA case and those in type-IIB case. This equation reveals that the anti-periodic standard Killing spinors may depend on the fiber coordinate ${\bar \chi}$. Second, $\epsilon$ transforms as a spinor. Even for the periodic Killing spinors, this fact does not immediately support our naive expectation. Although the standard periodic Killing spinors are constant ones in type-IIB coordinates, they may depend on $\bar{\chi}$ when they are written in terms of type-IIA coordinates. In order to fix the relationship between the type-IIA spinors and the type-IIB spinors, we need the spinor representation of the nonlinear diffeomorphism between $(\bar{x}_{4B}, \bar{x}_{5B})$ and $(\bar{x}_{4A}, \bar{x}_{5A})$. It is expected to be very involved to get an explicit expression of the spinor representation since the coordinate transformation is given in terms of the polar coordinates as in (\ref{new}).  

The resulting 4 Killing spinors survived Hopf-duality, depend only on the coordinate $x^+_A$, which is independent of the coordinate $\bar{\chi}$. Therefore no Killing spinor depending on the dual coordinate $\bar{\chi}$ survives Hopf-duality. It requires further analysis on the spinor transformation to understand the precise relationship between the Killing spinors in type-IIB and those in type-IIA under the Hopf T-duality. We may save this for a future work. 

We conclude this paper with some outlooks on future works.
Being a symmetry of the superstring theory, T-duality is believed to preserve supersymmetries in the context of full string theory. However at the level of low-energy effective supergravity, it may break supersymmetries. One good example is given by $AdS_5\times S^5$ case \cite{duff2} which arises as near horizon limit of D$3$-brane geometry. At the level of supergravity, Hopf-duality along the Hopf fiber direction over $CP^2$ breaks supersymmetries completely. Although there is no spin structure on $CP^2$, Hopf fibration of $S^5$ itself does not break supersymmetry because it is no more than renaming the coordinates. 
On $CP^2$ may we introduce a generalized spin structure 
to takes care of the Kaluza-Klein charges with which charged spinors are 
well-defined. Since Hopf-duality transforms these Kaluza-Klein gauge fields 
into the NS two-form fields, the supersymmetries in the dual theory can be 
seen only when string winding modes are included. We expect that the same is 
true for the system studied in this paper. 

It would be also interesting to understand the supernumerary supersymmetries 
in the context of boundary CFT. Recently the duality between 
superstrings in the plane wave background obtained from the near horizon 
geometry of D1/D5/F1/NS5-brane system and the orbifold CFT on the boundary was 
studied \cite{maldacena,lunin,strominger}. Since the system is S-dual to D1/D5 
system and their near-horizon geometries are the same, we expect that a similar 
supersymmetry enhancement may occur in the boundary orbifold CFT.


\section*{Acknowledgments}
We thank Sangmin Lee and Hyeonjoon Shin for stimulating discussions.
JHC (Project No. R01-2000-00021) and TL (Project No. R01-2000-00015) are supported in part by KOSEF. SN is supported by Korea Research Foundation Grant KRF-2001-041-D00049 and by BK21 Program of Korea Research Fund.

\begin{appendix}

\section{Killing spinor equations in type-IIB SUGRA}\label{a}

The fermionic fields in type-IIB supergravity are composed of the dilatino fields $\lambda$ and the gravitino fields $\psi_M$. Both fields are complex Weyl spinors. At the tree level, the unbroken supersymmetries are manifested by the the invariance of the above fermionic fields under the possible super transformations (See \cite{gsw} for details about this argument). In the vanishing fermion backgrounds, they are (in the Einstein frame)\cite{schwarz}
\begin{eqnarray}\label{killingiib}
&&\delta\lambda=i\Gamma^M\epsilon^*P_M-\frac{i}{24}\Gamma^{KLN}G_{KLN}\,\,\epsilon=0,\nonumber\\
&&\delta\psi_M={\cal D}_M\epsilon+\frac{1}{96}\left(\Gamma_M{}^{KLN}G_{KLN}-9\Gamma^{LN}G_{MLN}\right)\epsilon^*\nonumber\\
&&
\qquad+\frac{i}{4\cdot 480}\Gamma^{M_1\cdots M_5}F_{M_1\cdots M_5}\Gamma_M\epsilon =0,
\end{eqnarray}
where complex Weyl spinors satisfy $\Gamma^{(11)}\psi_M=\psi_M,$ $\Gamma^{(11)}\epsilon=\epsilon,$ and $\Gamma^{(11)}\lambda=-\lambda$. A few remarks are in order. Type-IIB supergravity, being an extended supergravity theory, possesses a noncompact global symmetry $SU(1,1)$ with $U(1)$ as its maximal compact subgroup. The scalar fields $V^\alpha{}_{\pm}$ of the theory parametrize the coset $SU(1,1)/U(1)$:
\begin{eqnarray}
\left(\matrix{V^1{}_-&V^1{}_+\cr
V^2{}_-&V^2{}_+
}
\right)=\frac{1}{2\sqrt{\tau_2}}
\left(\matrix{
\left(-\bar{\tau}+i\right)e^{i\nu}
&\left(-\tau+i\right)e^{-i\nu}\cr
\left(-\bar{\tau}-i\right)e^{i\nu}&\left(-\tau-i\right)
e^{-i\nu}
}
\right),
\end{eqnarray}
where the auxiliary real scalar field $\nu$ can be fixed to $\nu=0$ by the local $U(1)$ symmetry. The complex field $\tau$ is associated with the RR zero form field $C^{(0)}$ and the dilaton field $\phi$ as $\tau=C^{(0)}+ie^{-\phi}$. The field $P_M$ appearing in the above dilatino condition is $SU(1,1)$-invariant quantity constructed from $V^\alpha{}_{\pm}$:
\begin{eqnarray}
P_M=-\epsilon_{\alpha\beta}V^\alpha{}_+\partial_MV^\beta{}_+.
\end{eqnarray}
The $SU(1,1)$-invariant three form field $G$ in the gravitino condition is constructed as
\begin{eqnarray}
G=-\epsilon_{\alpha\beta}V^\alpha{}_+F^\beta=\frac{ie^{-i\nu}}{\sqrt{\tau_2}}\left(dC^{(2)}-\tau dB^{(2)}\right).
\end{eqnarray}
Here RR two-form field $C^{(2)}$ and NS two-form field $B^{(2)}$ constitute the $SU(1,1)$-doublet as
\begin{eqnarray}
&&F^1=-dC^{(2)}+idB^{(2)},\nonumber\\
&&F^2=\left(F^1\right)^*=-dC^{(2)}-idB^{(2)}.
\end{eqnarray} 

\section{Killing spinors in the $AdS_3\times S^3\times T^4$ background}\label{b}

In this section we give explicit forms of the Killing spinors in the $AdS_3\times S^3\times T^4$ background. In Ref. \cite{lu}, the Killing spinors in various $AdS_p\times S^q$ backgrounds are constructed in the horospherical coordinates. However in this paper, the Penrose limit is defined in terms of the global coordinates. Thus, it is more useful to write the Killing spinors in global coordinates. It would help us to understand the origin of supernumerary supersymmetries.

The orthonormal frame for the Einstein metric is composed of 
\begin{eqnarray}
\begin{array}{lll}
e^{(0)}=\zeta^{-\frac{1}{4}}R\cosh{\rho}\,dt,&\quad e^{(1)}=\zeta^{-\frac{1}{4}}R\cos{\theta}\,d\psi,&\quad e^{(2)}=\zeta^{-\frac{1}{4}}R\,d\rho,\\
e^{(3)}=\zeta^{-\frac{1}{4}}R\sinh{\rho}\,d\varphi,&\quad e^{(4)}=\zeta^{-\frac{1}{4}}R\,d\theta,&\quad e^{(5)}=\zeta^{-\frac{1}{4}}R\sin{\theta}\,d\chi,\\
e^{(i)}=\zeta^{-\frac{1}{4}}\zeta^{\frac{1}{2}}\,dx_i,&(i=6,\cdots, 9).&
\end{array}
\end{eqnarray}

The covariant derivatives are
\begin{eqnarray}
&&{\cal D}_t=\partial_t+\frac{1}{2}\sinh{\rho}\,\gamma_{(0)(2)},\quad {\cal D}_\rho=\partial_\rho,\quad {\cal D}_\varphi=\partial_\varphi-\frac{1}{2}\cosh{\rho}\,\gamma_{(2)(3)},\nonumber\\
&&{\cal D}_\psi=\partial_\psi-\frac{1}{2}\sin{\theta}\,\gamma_{(1)(4)},\quad{\cal D}_\theta=\partial_\theta,\quad {\cal D}_\chi=\partial_\chi-\frac{1}{2}\cos{\theta}\,\gamma_{(4)(5)},\nonumber\\
&&{\cal D}_i=\partial_i,\quad (i=6,\cdots, 9).
\end{eqnarray}

The gamma matrices in the coordinate frame and those in the orthonormal frame are related to each other as
\begin{eqnarray}
\begin{array}{lll}
\Gamma_t=R\,\zeta^{-\frac{1}{4}}\cosh{\rho}\,\gamma_{(0)},&\,\,\,
\Gamma_\psi=R\,\zeta^{-\frac{1}{4}}\cos{\theta}\,\gamma_{(1)},&\,\,\,
\Gamma_\rho=R\,\zeta^{-\frac{1}{4}}\,\gamma_{(2)},\\
\Gamma_\varphi=R\,\zeta^{-\frac{1}{4}}\sinh{\rho}\,\gamma_{(3)},&\,\,\,
\Gamma_\theta=R\,\zeta^{-\frac{1}{4}}\,\gamma_{(4)},&\,\,\,
\Gamma_\chi=R\,\zeta^{-\frac{1}{4}}\sin{\theta}\,\gamma_{(5)},\\
\Gamma_i=\zeta^{\frac{1}{4}}\,\gamma_{(i)},\quad(i=6,\cdots, 9).&
\end{array}
\end{eqnarray}
and
\begin{eqnarray}\label{gammaup}
\begin{array}{lll}
\Gamma^t=R^{-1}\,\zeta^{\frac{1}{4}}\cosh^{-1}{\rho}\,\gamma^{(0)},&\,\,
\Gamma^\psi=R^{-1}\,\zeta^{\frac{1}{4}}\cos^{-1}{\theta}\,\gamma^{(1)},&\,\,
\Gamma^\rho=R^{-1}\,\zeta^{\frac{1}{4}}\,\gamma^{(2)},\\
\Gamma^\varphi=R^{-1}\,\zeta^{\frac{1}{4}}\sinh^{-1}{\rho}\,\gamma^{(3)},&\,\,
\Gamma^\theta=R^{-1}\,\zeta^{\frac{1}{4}}\,\gamma^{(4)},&\,\,
\Gamma^\chi=R^{-1}\,\zeta^{\frac{1}{4}}\sin^{-1}{\theta}\,\gamma^{(5)},\nonumber\\
\Gamma^i=\zeta^{-\frac{1}{4}}\,\gamma^{(i)},\quad (i=6,\cdots, 9).&
\end{array}\\
\end{eqnarray}

The $SU(1,1)$-invariant three-form field $G$ is
\begin{eqnarray}
G=2i\zeta^{\frac{1}{2}}g\,\alpha'N_5\left(\cosh{\rho}\sinh{\rho}\,dt\wedge d\rho\wedge d\varphi+\cos{\theta}\sin{\theta}\,d\psi\wedge d\theta\wedge d\chi\right).
\end{eqnarray}

In the vanishing fermion background, the dilatino condition is 
\begin{eqnarray}\label{adsdilatino}
\delta\lambda&=&\frac{1}{2}\zeta^{\frac{1}{2}}g\,\alpha'N_5\left(\Gamma^{t\rho\varphi}\cosh{\rho}\sinh{\rho}+\Gamma^{\psi\theta\chi}\cos{\theta}\sin{\theta}\right)\epsilon\nonumber\\
&=&\frac{1}{2}R^{-1}\zeta^{\frac{1}{4}}\left(\gamma^{(0)(2)(3)}+\gamma^{(1)(4)(5)}\right)\epsilon=0.
\end{eqnarray}

The gravitino conditions are
\begin{eqnarray}\label{adsgravitino}
&&\delta\psi_t=\partial_t\epsilon-\frac{1}{2}\sinh{\rho}\,\gamma^{(0)(2)}\epsilon-\frac{i}{8}\cosh{\rho}\left(3\gamma^{(2)(3)}+\gamma^{(0)(1)(4)(5)}\right)\epsilon^*=0,\nonumber\\
&&\delta\psi_\psi=\partial_\psi\epsilon-\frac{1}{2}\sin{\theta}\gamma^{(1)(4)}\,\epsilon-\frac{i}{8}\cos{\theta}\left(3\gamma^{(4)(5)}+\gamma^{(0)(1)(2)(3)}\right)\epsilon^*=0,\nonumber\\
&&\delta\psi_\rho=\partial_\rho\epsilon+\frac{i}{8}\left(3\gamma^{(0)(3)}-\gamma^{(1)(2)(4)(5)}\right)\epsilon^*=0,\nonumber\\
&&\delta\psi_\varphi=\partial_\varphi\epsilon-\frac{1}{2}\cosh{\rho}\,\gamma^{(2)(3)}\epsilon-\frac{i}{8}\sinh{\rho}\left(3\gamma^{(0)(2)}
+\gamma^{(1)(3)(4)(5)}\right)\epsilon^*=0,\nonumber\\
&&\delta\psi_\theta=\partial_\theta\epsilon+\frac{i}{8}\left(3\gamma^{(1)(5)}-\gamma^{(0)(2)(3)(4)}\right)\epsilon^*=0,\nonumber\\
&&\delta\psi_\chi=\partial_\chi\epsilon-\frac{1}{2}\cos{\theta}\gamma^{(4)(5)}\,\epsilon-\frac{i}{8}\sin{\theta}\left(3\gamma^{(1)(4)}+\gamma^{(0)(2)(3)(5)}\right)\epsilon^*=0,\nonumber\\
&&\delta\psi_i=\partial_i\epsilon=0, \quad (i=6,\cdots 9).
\end{eqnarray}
The last equation tells us that the Killing spinors are constant around the torus, $T^4$.
With the dilatino condition,
\begin{eqnarray}
\left(\gamma^{(0)(1)(2)(3)(4)(5)}+1\right)\epsilon=0,
\end{eqnarray} 
the first six gravitino conditions become simplified as
\begin{eqnarray}
&&\partial_\mu\left(\matrix{\epsilon_1\cr \epsilon_2}\right)=\frac{1}{2}\Omega_\mu\left(\matrix{\epsilon_1\cr \epsilon_2}\right),\nonumber\\
\mbox{with}&&\qquad \Omega_t= \sinh{\rho}\,\gamma^{(0)(2)}\otimes 1\!\!1 + \cosh{\rho}\,\gamma^{(2)(3)}\otimes\sigma_1,\nonumber\\
&&\qquad \Omega_\rho= -\gamma^{(0)(3)}\otimes\sigma_1,\nonumber\\
&&\qquad \Omega_\varphi= \cosh{\rho}\,\gamma^{(2)(3)}\otimes 1\!\!1 + \sinh{\rho}\,\gamma^{(0)(2)}\otimes\sigma_1,\nonumber\\
&&\qquad \Omega_\psi= \sin{\theta}\,\gamma^{(1)(4)}\otimes 1\!\!1 + \cos{\theta}\,\gamma^{(4)(5)}\otimes\sigma_1,\nonumber\\
&&\qquad \Omega_\theta=  -\gamma^{(1)(5)}\otimes\sigma_1,\nonumber\\
&&\qquad \Omega_\chi= \cos{\theta}\,\gamma^{(4)(5)}\otimes 1\!\!1 + \sin{\theta}\,\gamma^{(1)(4)}\otimes\sigma_1,
\end{eqnarray}
where we note that $\Omega_\mu^2=-1\!\!1_{64}$ except $\Omega_\rho^2=1\!\!1_{64}$, and $\Omega_\varphi=\left(1\!\!1_{32}\otimes\sigma_1\right)\Omega_t$ and $\Omega_\chi=\left(1\!\!1_{32}\otimes\sigma_1\right)\Omega_\psi$. Making use of these facts and the following identities,
\begin{eqnarray}
e^{-\frac{\rho}{2}\Omega_\rho}\Omega_t e^{\frac{\rho}{2}\Omega_\rho}=\gamma^{(2)(3)}\otimes\sigma_1,\qquad e^{-\frac{\theta}{2}\Omega_\theta}\Omega_\psi e^{\frac{\theta}{2}\Omega_\theta}=\gamma^{(4)(5)}\otimes\sigma_1,
\end{eqnarray}
one can easily obtain the following Killing spinors:
\begin{eqnarray}\label{adsKilling}
\epsilon&=&e^{-\frac{\rho}{2}\gamma^{(0)(3)}\otimes\sigma_1}e^{\frac{t}{2}\gamma^{(2)(3)}\otimes\sigma_1+\frac{\varphi}{2}\gamma^{(2)(3)}\otimes1\!\!1}\cdot\nonumber\\
&&\qquad \cdot e^{-\frac{\theta}{2}\gamma^{(1)(5)}\otimes\sigma_1}e^{\frac{\psi}{2}\gamma^{(4)(5)}\otimes\sigma_1+\frac{\chi}{2}\gamma^{(4)(5)}\otimes1\!\!1}\epsilon_0.
\end{eqnarray}
The constant spinors $\epsilon_0$ are constrained only by the dilatino condition;
\begin{eqnarray}
\left(\gamma^{(0)(2)(3)}+\gamma^{(1)(4)(5)}\right)\epsilon_0=0.
\end{eqnarray}
Hence there are 16 unbroken supersymmetries which are anti-periodic for the cycles of the global coordinates,
\begin{eqnarray}
\varphi \sim \varphi+2\pi,\qquad \psi \sim \psi+2\pi,\qquad \chi \sim \chi+2\pi.\end{eqnarray}

\section{Killing spinor equations in type-IIA SUGRA}\label{iia}

In the vanishing fermion background of type-IIA supergravity, the dilatino condition and the gravitino conditions for the unbroken supersymmetries are
\begin{eqnarray}
\sqrt{2}\delta\lambda&=&-\frac{1}{2}{\cal D}_M\phi'\,\Gamma^M\Gamma^{11}\epsilon+\frac{1}{24}e^{-\frac{\phi'}{2}}H^{(3)}_{M_1M_2M_3}\Gamma^{M_1M_2M_3}\epsilon\nonumber\\
&&-\frac{3}{8\cdot 2!}e^{\frac{3\phi'}{4}}F^{(2)}_{M_1M_2}\Gamma^{M_1M_2}\epsilon+\frac{1}{8\cdot 4!}e^{\frac{\phi'}{4}}F^{(4)}_{M_1M_2M_3M_4}\Gamma^{M_1M_2M_3M_4}\Gamma^{11}\epsilon=0,\nonumber\\
\delta\psi_M&=&{\cal D}_M\epsilon+\frac{1}{96}e^{-\frac{\phi'}{2}}H^{(3)}_{M_1M_2M_3}\left(\Gamma_M{}^{M_1M_2M_3}-9\delta_M^{M_1}\Gamma^{M_2M_3}\right)\Gamma^{11}\epsilon\nonumber\\
&&-\frac{1}{64}e^{\frac{3\phi'}{4}}F^{(2)}_{M_1M_2}\left(\Gamma_M{}^{M_1M_2}-14\delta_M^{M_1}\Gamma^{M_2}\right)\Gamma^{11}\epsilon\nonumber\\
&&+\frac{1}{256}e^{\frac{\phi'}{4}}F^{(4)}_{M_1M_2M_3M_4}\left(\Gamma_M{}^{M_1M_2M_3M_4}-\frac{20}{3}\delta_M^{M_1}\Gamma^{M_2M_3M_4}\right)\epsilon=0,
\end{eqnarray}
where the spinor $\epsilon$ is of Majorana but not of Weyl. Here $\phi'$ is the dilaton field of type-IIA supergravity and the three-form field, $H^{(3)}=dB^{(2)}$ is the NS field strength.

\section{Hopf-fibration}\label{hopf}

We consider a 3-sphere and construct an explicit form of the Hopf-fibration. A 3-sphere is described by the following metric:
\begin{eqnarray}
d\Omega_3^2&=&\cos^2{\theta}\,\,d\psi^2+d\theta^2+\sin^2{\theta}\,\,d\chi^2,\nonumber\\
&&\quad 0\le\theta\le\frac{\pi}{2},\quad \psi\sim\psi+2\pi,\quad \chi\sim\chi+2\pi.
\end{eqnarray}
Here the character $\Omega$ describing the sphere metric is not to be confused with the operators $\Omega_\mu$ appearing in the Killing spinor equations.
Introducing new coordinates as
\begin{eqnarray}\label{new}
\theta&=&\frac{1}{2}\left(\bar{\theta}+\frac{\pi}{2}\right),\quad \psi=\frac{1}{2}\left(\bar{\psi}+\bar{\chi}\right),\quad \chi=\frac{1}{2}\left(\bar{\psi}-\bar{\chi}\right),\nonumber\\
&&\quad -\frac{\pi}{2}\le \bar{\theta}\le \frac{\pi}{2},\quad \bar{\psi}\sim\bar{\psi}+2\pi,\quad \bar{\chi}\sim \bar{\chi}+4\pi,
\end{eqnarray}
we get the following Kaluza-Klein type metric:
\begin{eqnarray}\label{fiber}
d\Omega_3^2&=&\frac{1}{4}\left(\cos^2{\bar{\theta}}\,\,d\bar{\psi}^2+d\bar{\theta}^2\right)+\frac{1}{4}\left(d\bar{\chi}-\sin{\bar{\theta}}\,\,d\bar{\psi}\right)^2\nonumber\\
&=&\frac{1}{4}\left[d\bar{\Omega}_2^2+\left(d\bar{\chi}+\bar{\cal A}\right)^2\right].
\end{eqnarray}
Therefore 3-sphere can be represented as $U(1)$-fibered 2-sphere, where the Kaluza-Klein field strength is determined as $\bar{\cal F}=d\bar{\cal A}=-d\bar{\Omega}_2$.
\end{appendix}


\begin{thebibliography}{99}

\bibitem{maldacena} D. Berenstein, J. Maldacena, and H. Nastase, ``Strings in flat space and pp waves from ${\cal N}=4$ super Yang Mills,'' JHEP {\bf 0204} (2002) 013 [hep-th/0202021].

\bibitem{kowalski} J. Kowalski-Glikman, ``Vacuum states in supersymmetric Kaluza-Klein Theory, \pl{134}{1984}{194}.

\bibitem{penrose}
R. Penrose, ``Any space-time has a plane wave as a limit" in M. Cahen and M. Flato, ed. {\it Differential geometry and relativity}, page 271, D. Reidel, Dordrecht, 1976.

\bibitem{gueven1}
R.~G\"{u}ven,
``Plane Waves In Effective Field Theories Of Superstrings,'' \pl{191}{1987}{275}.

\bibitem{gueven2} R.~G\"{u}ven, ``Plane Wave Limits and T-Duality,'' \pl{482}{2000}{255} [hep-th/0005061].

\bibitem{blau2} M. Blau, J. Figueroa-O'Farrill, C. Hull, and G. Papadopoulos, ``Penrose limits and maximal supersymmetry,'' \cqg{19}{2002}{L87} [hep-th/0201081].

\bibitem{blau3} M. Blau, J. Figueroa-O'Farrill, and G. Papadopoulos, ``Penrose limits, supergravity and brane dynamics,'' hep-th/0202111.

\bibitem{blau1} M. Blau, J. Figueroa-O'Farrill, C. Hull, and G. Papadopoulos, ``A new maximally supersymmetric background of IIB superstring theory,'' JHEP {\bf 0201} (2002) 047 [hep-th/0110242].

\bibitem{figueroa} J. Figueroa-O'Farrill and G. Papadopoulos, ``Homogeneous fluxes, branes and a maximally supersymmetric solution of M-theory,'' JHEP {\bf 0108} (2001) 036 [hep-th/0105308].

\bibitem{cvetic} M. Cvetic, H. L\"{u}, and C.N. Pope, ``Penrose Limits, PP-Waves and Deformed M2-branes,'' hep-th/0203082.

\bibitem{gauntlett}  J.P. Gauntlett and C.M. Hull, ``pp-waves in 11-dimensions with Extra Supersymmetry,'' JHEP {\bf 0206} (2002) 013 [hep-th/0203255].

\bibitem{lu2} H. L\"{u} and J.F. V\'{a}zquez-Poritz, ``Penrose Limits of Nonstandard Brane Interactions,'' hep-th/0204001.

\bibitem{singh} H. Singh, ``M5-branes with 3/8 Supersymmetry in pp-wave Background,'' hep-th/0205020.

\bibitem{kumar} M. Alishahiha and A. Kumar, ``D-brane Solutions from New Isometries of PP-waves,'' hep-th/0205134.

\bibitem{duff2} M.J. Duff, H. Lu, and C.N. Pope, ``$AdS_5\times S^5$ Untwisted,'' \np{532}{1998}{181} [hep-th/9803061].

\bibitem{duff}  M.J. Duff, H. Lu, and C.N. Pope, ``$AdS_3 \times S^3$ (Un)twisted and squashed, and an O(2,2;Z) multiplet of Dyonic strings,'' \np{544}{1999}{145} [hep-th/9807173].

\bibitem{bergshoeff} E. Bergshoeff, R. Kallosh, and T. Ort\'{i}n, ``Duality versus Supersymmetry and Compactification,'' \pl{343}{1995}{103} [hep-th/9410230].


\bibitem{schwarz} J.H. Schwarz, ``Covariant Field Equations of Chiral N=2 D = 10 Supergravity,'' \np{226}{1983}{269}.

\bibitem{gsw} M.B. Green, J.H. Schwarz, and E. Witten, {\it Superstring Theory}, Vol. 2, (Cambridge University Press, 1987).


\bibitem{huq}
M.~Huq and M.~A.~Namazie,
``Kaluza-Klein Supergravity In Ten-Dimensions,'' \cqg{2}{1985}{293},
[Erratum-ibid.\  {\bf 2} (1985) 597].

\bibitem{giani} F. Giani and M. Pernici, ``$N=2$ Supergravity in ten dimensions,'' \pr{30}{1984}{325}.

\bibitem{hassan} S.F. Hassan, ``T-Duality, space-time spinors and R-R fields in curved backgrounds,'' \np{568}{2000}{145} [hep-th/9907152].

\bibitem{lunin} O. Lunin and S. Mathur, ``Rotating deformations of $AdS_3\times S^3$, the orbifold CFT and strings in the pp-wave limit,'' hep-th/0206107.

\bibitem{strominger} J.~Gomis, L.~Motl, and A.~Strominger, ``pp-wave / CFT(2) duality,'' hep-th/0206166.

\bibitem{lu} H. L\"{u}, C.N. Pope, and J. Rahmfeld, ``A Construction of Killing Spinors on $S^n$,'' \jmp{40}{1999}{4518} [hep-th/9805151].


\end{thebibliography}
\end{document}